\documentclass[preprint]{aastex}
\usepackage{amssymb,times}

\shorttitle{Black Hole Remnant in SN 1979C}

\begin{document}

\title{Evidence for a Possible Black Hole Remnant in the Type IIL 
Supernova 1979C}

\author{D.~J. Patnaude, A.~Loeb, \& C.~Jones}
\affil{Harvard-Smithsonian Center for Astrophysics, 60 Garden Street, 
Cambridge, MA 02138, USA}

\begin{abstract}

We present an analysis of archival X-ray observations of the Type IIL 
supernova SN~1979C. We find that its X-ray luminosity is remarkably 
constant at ($6.5\pm0.1$) $\times$ 10$^{38}$ erg s$^{-1}$ over
a period of 12 years between 1995 and 2007. The high and steady
luminosity is considered as possible 
evidence for a stellar-mass ($\sim$ 5--10M$_{\odot}$) 
black hole accreting material from either a supernova fallback disk 
or from a binary companion, or possibly from emission
from a central pulsar wind nebula.  We find that the bright and steady
X-ray light curve is not consistent with either a model for a supernova
powered by magnetic braking of a rapidly rotating magnetar, 
or a model where the blast wave is
expanding into a dense circumstellar wind.

\end{abstract}


\section{Introduction}

\citet{kasen09} and \citet{woosley09} have recently suggested that a
class of Type IIL supernovae (SNe) may be powered by the birth of a
magnetar during the SN event. These SNe include SN~1961F, SN~1979C,
SN~1980K, and SN~1985L. In particular, \citet{kasen09} argue that for
a supernova with ejecta mass $M_{\rm ej} = 5M_{\odot}$ that forms a
magnetar with an initial period of $P_i=10$ ms, the supernova can reach
peak luminosities consistent with what was observed in these
events. They also show that even brighter (M$_B$ = -21) events can
occur, such as the ultra-bright Type IIL SNe 2005ap and 2008es.

Motivated by \citet{kasen09} and \citet{woosley09}, we compared the
time evolution of the X-ray luminosity of SN~1979C to the form
expected from a magnetar-powered SN. In \S~2, we discuss SN 1979C and
examine its X-ray luminosity. We find that contrary to the expectation
of these papers, the evolution of this SN is not consistent with the
magnetar model. Additionally, we show that the X-ray emission spectrum
exhibits evidence for hard X-ray emission, possibly originating from
a central source. We show that the X-ray luminosity has been steady over
the lifetime of the SN evolution and argue that this is due to an accreting
stellar-mass black hole remnant at the center of SN~1979C.

\section{SN~1979C}

The Type IIL SN~1979C, discovered on April 19 1979 by G.~Johnson
\citep{mattei79} and its host galaxy NGC 4321 (M100, at a distance 
of $15.2\pm0.1$ Mpc; \citealt{freedman01}) have been extensively
observed in the radio \citep{weiler86,montes00,bartel03,bartel08,marcaide09}, 
optical \citep{fesen93,milisavljevic09} and X-ray
\citep{immler98,kaaret01,immler05} bands. At optical wavelengths, the
H$\alpha$ flux has decreased by $\sim$ 35\% between 1993 and 2008,
while forbidden line emission from lines such as [\ion{O}{3}] has
increased $\sim$ 50\% \citep[see Figure~3 of][]{milisavljevic09}. 
The increase in the
forbidden line emission is evidence of shock heating of the ejecta by
the reverse shock.  Similar to the H$\alpha$ emission, the radio
emission also shows a decline of a factor of $\sim$ 5 over the lifetime
of the SN \citep{bartel08}, though early
observations provided evidence for quasi-periodic oscillations that may be
the result of a modulation in the progenitor's circumstellar environment
by a binary companion \citep{weiler92, schwarz96}. 

Recently, \citet{bartel08} noted that the in conjunction with 
a decrease in the radio luminosity the spectrum has begun to flatten. 
They note that the
flattening spectrum would be expected for a supernova with emission
from a compact remnant that is beginning to appear in the shell as
the expanding remnant becomes increasingly transparent. They point out, 
however, that the flattened spectrum could also be due to synchrotron 
emission from the shell as electrons are accelerated to GeV energies.
We note that in
this scenario, it would be reasonable to expect an increasing
hard X-ray flux as electrons are accelerated to TeV energies.
More recently, \citet{marcaide09} presented a revised analysis of
the radio expansion of SN~1979C and found that the expanding
blastwave is in near free expansion with $m$ = 0.91$\pm$0.09 for
$R \propto t^m$. This seems contrary to X-ray observations which
suggest that the blastwave has been strongly decelerated by 
a dense circumstellar interaction \citep{immler05}. While the
optical and radio emission have showed evolution over $\sim$ 25 years,
X-ray
observations of SN~1979C show a markedly different behavior. As seen in
Table~\ref{tab:79c}, the X-ray luminosity has remained remarkably
constant over its 30 yr lifetime.

\citet{kasen09} suggest that Type IIL SNe such as 
SN~1979C may have been magnetar
powered. In the context of a magnetic dipole model, one can write the
spindown luminosity as $L_p = 
5 \times 10^{42} B_{14}^2 P_{i}^{-4}$ erg s$^{-1}$ 
for a magnetar with magnetic field B$_{14}$ (in
units of 10$^{14}$ Gauss) and period $P_{i}$ (in units of 10 ms), or as

\begin{equation}
L_p = 2 \times 10^{42} B_{14}^2 \left({t}/{\rm yr}\right)^{-2}~{\rm erg~s^{-1}} .
\end{equation}

\begin{figure}
\includegraphics[width=0.5\textwidth]{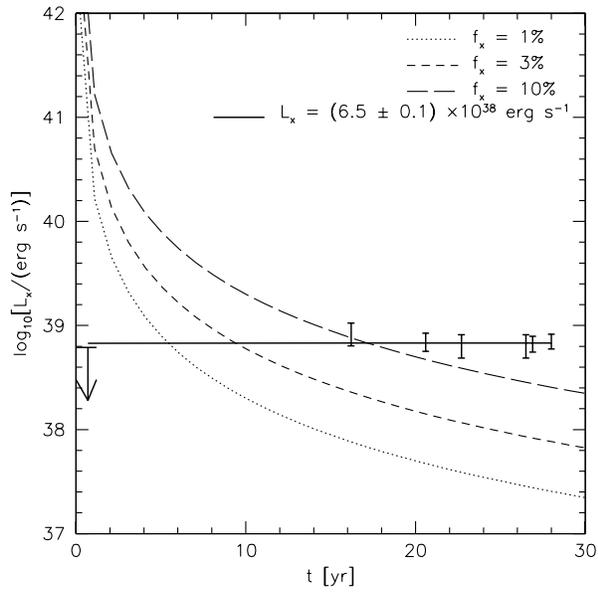}
\caption{Evolution of the X-ray luminosity of SN~1979C. The dotted, 
dashed, and long-dashed diagonal lines correspond to the
evolution of the SN luminosity if it were powered by the magnetic
dipole spindown of a central magnetar for various X-ray efficiencies,
$f_x$. The data points correspond to the 0.3--2.0 keV X-ray luminosities
with 1-$\sigma$ uncertainties, assuming a distance of 15.2
Mpc. The upper limit from the {\it Einstein} observatory is depicted
as an arrow. The heavy solid line corresponds to our best fit value of
a nearly constant X-ray luminosity.}
\label{fig:x1979c}
\end{figure}

The observed X-ray luminosity, $L_{X}$, can be expressed as some
fraction, $f_x$, of the magnetar spindown luminosity, i.e. $L_X= f_x
L_p$. In Figure~\ref{fig:x1979c} we plot the evolution of the X-ray
luminosity for a dipole magnetar-powered SN assuming that 1, 3, or 10\%
of the spindown power is transferred into X-rays, with $B_{14}=1$ and
$P_{i}$ = 1.  We also indicate the X-ray luminosity (from
Table~\ref{tab:79c}) as observed by the {\em ROSAT}, {\em XMM--Newton},
{\em Chandra}, and {\em Swift} observatories, 
as well as the early upper limit from
the {\em Einstein} observatory HRI, where the model predicted X-ray 
luminosity is more than two orders of magnitude above the 
{\em Einstein} upper limit. As seen in Figure~\ref{fig:x1979c} and
also in Table~\ref{tab:79c}, the observed X-ray luminosity\footnote{to 
convert from the observed count rates to source fluxes and luminosities, we 
assume an absorbed thermal bremsstrahlung model with 
n$_{\mathrm{H}}$ = 2.4$\times$10$^{20}$ cm$^{-2}$ and $kT$ = 0.5 keV.} 
is remarkably
constant over time, with L$_X$ = (6.5$\pm$0.1)$\times$10$^{38}$ erg
s$^{-1}$. 

Most importantly, even if the magnetar powered the
SN only at early times, it would have likely resulted in a clear
detection by the {\em Einstein} observatory.
The above equation applies to a magnetic dipole spindown. More
generally, the magnetar luminosity can be written as

\begin{equation}
L_p \propto \frac{(l-1)}{(1+t/t_p)^l},
\end{equation}

where $t_p$ is the initial spindown time and $l=2$ for magnetic dipole
spindown. At late times, $L_p\propto
t^{-l}$, and for all physical values of $l$ the luminosity
decreases with time. If the X-ray luminosity is some fraction of the
spindown luminosity, the fraction will need to increase with time,
potentially up to unphysical values greater than unity, in order for
the observed X-ray luminosity to remain constant.

\subsection{Spectral Modeling}

\begin{figure}
\includegraphics[width=0.5\textwidth]{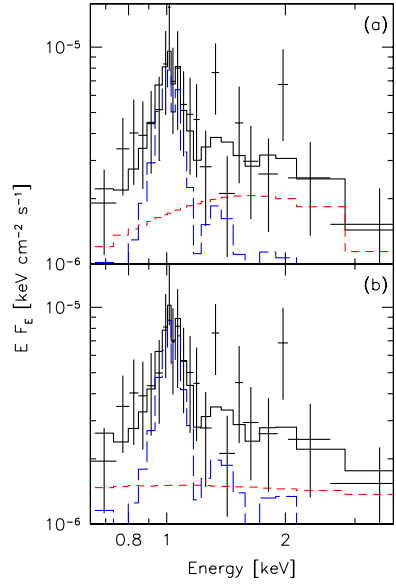}
\caption{The unfolded {\em Chandra} ACIS-S spectrum of SN~1979C.
Panel {\em (a)} shows the data with its 1-$\sigma$ error bars
(crosses), which is fitted as a combination of a thermal plasma
component (blue long-dashed line), and emission from a relativistic
accretion disk around a Kerr black hole (red short-dashed line). For
the thermal component, we infer a temperature of $kT$ = 1.2 keV, and
for the disk component, we infer a black hole mass of
$M_{\mathrm{BH}}=5.2M_{\sun}$. Panel {\em (b)} shows the same thermal
plasma model as in panel {\em (a)} together with a power-law fit (red
short-dashed line) with a spectral slope for the flux per unit energy
$F_E \propto E^{-\Gamma}$ of $\Gamma$ = 2.1. In both panels, the fitted
absorption is consistent with Galactic foreground of a hydrogen column
density, $N_{\mathrm{H}}$ = 2.5 $\times$ 10$^{20}$ cm$^{-2}$.}
\label{fig:79c_acis}
\end{figure}

The combined {\em Chandra} ACIS-S spectrum of SN~1979C is shown in
Figure~\ref{fig:79c_acis}. \citet{immler05} fit the {\em XMM-Newton}
observation to a two component thermal plasma model (MEKAL model in
{\sc Xspec}\footnote{http://heasarc.nasa.gov/xanadu/xspec/}), 
with $kT_{low}$ = 0.77$_{-0.19}^{+0.17}$ keV and
$kT_{high}$ = 2.31$_{-0.66}^{+1.95}$ keV. They did not find any
evidence for strong absorption, with a fitted hydrogen column density
$N_H \approx$ 2.5 $\times$ 10$^{20}$ cm$^{-2}$ \citep[see also
][]{dickey90}, and attribute the low temperature component to
recently shocked ejecta, and the high temperature component to
emission from swept--up circumstellar material (CSM). In their model,
the shocked circumstellar emission arises from a dense wind, with
densities of $\la$ 10$^{5}$ cm$^{-3}$ out to radii of 0.13 pc,
yielding column densities $\la 10^{22}~{\rm cm^{-2}}$. 
\citet{immler05} argue that VLBI imaging at $t$ = 22 yr show no
evidence for clumping in the CSM, indicating that the CSM is
remarkably uniform. If this were the case, then both the {\em
XMM--Newton} and {\em Chandra} ACIS-S spectrum would show strong
absorption at low energies.  We fitted the {\em Chandra} ACIS-S
spectrum to a two component thermal plasma model, and found that the
spectrum can be described by two thermal plasmas with temperatures
$kT =0.75_{-0.3}^{+0.2}$ and 1.86$_{-0.72}^{+0.88}$ keV, which 
shows no evidence for the level of absorption
expected from the above properties of a hypothetical dense smooth CSM.
We note that our fitted parameters and those of \citet{immler05}
are consistent.

Recently, \citet{bietenholz10} reported on VLBI detection of a central
source in SN~1986J. They argue that the small proper motion of the
source (500 $\pm$ 1500 km s$^{-1}$) provides evidence that the source
is either a central neutron star or black hole, but they do point out
that it might also be a dense CSM condensate seen
along the line of sight, as similar clumped regions appear in the
northeastern region of the SN~1986J shell. Additionally, as 
previously pointed out, \citet{marcaide09} showed that the 
blastwave of SN~1979C is still in free expansion, and \citet{bartel08} 
argued that the flattening radio spectrum could be evidence for
emission from a central compact remnant. We thus chose to fit the
X-ray spectrum of SN~1979C as a combination of shocked plasma, either
arising from shocked CSM or ejecta, along with a component associated
with a central source. The resultant fits are shown in
Figure~\ref{fig:79c_acis}. In both cases, we found that the thermal
component is well described by a plasma with $kT$ =
1.1$_{-0.12}^{+0.14}$ keV in collisional ionization equilibrium. We
first modeled the hard emission as a power-law (bottom panel of
Fig.~\ref{fig:79c_acis}), and found a power-law index for the flux per
unit energy $\Gamma$ = 2.2$_{-0.4}^{+0.3}$.  We also fit the data to a
model for an accreting central black hole, using the \texttt{kerrbb}
model in {\sc Xspec}, shown in the top panel of Figure~\ref{fig:79c_acis}.
We found that the hard component of the spectrum can be well fit for a
black hole mass of $\sim (5.2 \pm 0.8) M_{\odot}$. We do note, however,
that while the X-ray spectrum is well described by a spectrum from an
accreting black hole, we do concede that a powerlaw fit such as that
shown in the bottom panel of Figure~\ref{fig:79c_acis} is still consistent
with emission from either a central compact source such as a pulsar
wind nebula, or from synchrotron emission arising from shock
accelerated electrons.

\subsection{A Black Hole Remnant?}

While a two component thermal model can describe the X-ray spectrum
for SN~1979C, with the hard component having a $kT$ $\ga$ 2 keV, 
as seen in Figure~\ref{fig:79c_acis}, the data are equally well described by a
thermal component, either from shocked CSM or shocked ejecta, and a
hard spectral component, possibly from a central source.
\citet{immler05} modeled the X-ray emission as arising from the
expansion of the blastwave into a dense circumstellar wind, with
values ranging from 10$^{4}$ -- 10$^{7}$ cm$^{-3}$ out to radii of
4$\times$10$^{17}$ cm.  Ahead of the blastwave, the circumstellar
density in their model is still 10$^{3}$ cm$^{-3}$ at a distance of 1
pc \citep[c.f. Figure~6 of ][]{immler05}, which would add an absorbing
column density of $\sim 3\times 10^{21}$ cm$^{-2}$, well in excess of
the measured absorption in the spectrum of SN~1979C.

Additionally, in models for supernova remnant (SNR) evolution where
the blast wave is expanding into a wind, the X-ray (and optical and
radio) emissivity {\it from the expanding blast wave} should decrease
with increasing blast wave radius. This is because the blast wave
encounters the densest material at small radii, and sweeps over less
dense material as it expands to a larger volume.  At the late times of
interest here, the ejecta is optically thin \citep{Chev}.  The
free-free emission measure scales quadratically with the particle
number density $n$ and linearly with the volume $V\propto r^3$, so
that for the typical $n\propto r^{-2}$ radial profile of a wind,
$L_X\propto n^{2} V \propto r^{-1} $ decreases with time.  Other SN
remnants (SNRs) do show evidence for a strong interaction with a
circumstellar medium 
\citep[{\it c.f.,} SN~1994I and SN~1993J;][]{immler02,chandra09}. 
SN~1993J was observed
with {\em Chandra} in 2000 and again in 2008. Analysis of these data
show that the X-ray luminosity has dropped from $\sim 6\times
10^{38}~{\rm erg~s^{-1}}$ in 2000 to $\sim 1.5\times 10^{38}~{\rm
erg~s^{-1}}$ in 2008. This sharp drop in X-ray luminosity is
inconsistent with the near steady X-ray emission from SN~1979C.

A plausible explanation for the constant
luminosity would be to associate it with Eddington-limited
accretion onto a central compact object of mass $M_x$. Based on the
Eddington luminosity value,

\begin{equation}
L_{\rm Edd} = 1.4\times 10^{38}\left(\frac{M_x}{M_{\odot}}\right) 
~{\rm erg~s^{-1}} ,
\end{equation}

with a modest bolometric correction \citep[{\it e.g.} L$_{H\alpha}$ =
  1.6$\times$10$^{37}$, L$_{UV}$ = 9$\times$10$^{36}$ and
  L$_{1.6{\mathrm GHz}}$ = 1.6$\times$10$^{36}$ erg
  s$^{-1}$;][]{immler05,bartel03}, this implies that the accreting
object has a mass of $5$--$10 M_\odot$, which is within the typical
range associated with stellar-mass black holes \citep{Mc}.  The hard
component of the X-ray spectrum of the source shown in
Figure~\ref{fig:79c_acis} is consistent with the spectrum of an
Eddington-limited black hole in an X-ray binary such as LMC X-3
\citep[see Fig.~1 in ][]{Davis06}, and our spectral fits to the data
imply a black hole mass of $\ga$ 5M$_{\odot}$.  It is plausible to
expect that a black hole forms in a type IIL SNe. According to
Figure~2 of \citet{heger03}, solar metallicity stars with progenitor
masses $\simeq$ 25$M_{\odot}$ can follow a track that leads to black
hole formation. \citet{heger03} also point out that 25\% of low mass
(M $\sim$ 20M$_{\odot}$) progenitors with solar metallicity will
produce black hole remnants by fallback. Since $\sim$ 3\% of
core-collapse SNe are Type IIL \citep{smartt09}, 
this implies that some
Type IIL's will form a black hole remnant, consistent with predictions
for theoretical black hole mass distributions from core-collapse SNe
\citep{fryer01}.  Interestingly, \citet{heger03} also suggest that
Type IIL/b SNe are produced in binaries, which would avoid the need
for a fallback accretion disk in the SN \citep{P00,Wang06} as the
binary companion would provide the accreting material. There has
been some evidence for a binary companion to the progenitor of
SN~1979C, seen as a modulation of the early time radio light curve.

The existence of an accreting black hole remnant at the center of 
SN~1979C is consistent with the upper limit on the early X-ray luminosity
established by the {\em Einstein} Observatory. 
For supernova models with
2--5M$_{\odot}$ of ejecta and explosion energies of 2$\times$10$^{51}$ ergs,
the expanding ejecta bubble becomes optically thin to keV X-rays on a 
timescale of 15 yrs. The upper limit on the {\em Einstein} observation
may reflect that early phase of the SNR evolution when the ejecta 
is optically thick or the time delay required
to establish a stable accretion flow around the black hole remnant, which
is calculated to be about twice the recombination timescale, 
$\sim$ 1 yr \citep{zampieri98,balberg00}. 

\citet{milisavljevic09} note the presence of a 
Wolf--Rayet ``bump'' in the optical emission spectrum of SN~1979C which
may be associated with the progenitor. They mention that 
several metal--rich \ion{H}{2} regions in M100 have been observed to possess
WR stars \citep{vandyk99,pindao02}, and {\em HST} imaging shows several young
blue stars of age 4--6 Myr in the vicinity of SN~1979C. Photometry of 
the clusters' stellar population yields an estimate of the progenitor's
mass to be (18)$\pm$3 M$_{\odot}$. This is consistent with the expected
progenitors for Type IIL SN and with the required progenitor mass to
form a black hole during the SN explosion, possibly from a fallback disk.

We have also analyzed {\em Chandra}
and {\em ROSAT} observations of other two other Type IIL SNe,
SN~1980K and SN~1985L. We have found that the X-ray luminosity of
SN~1979C is more than an order of magnitude larger than that of SN
1980K (L$_X$ $\approx$ 5.5$\times$10$^{37}$ erg s$^{-1}$) and SN~1985L
(L$_X$ $\lesssim$ 4$\times$ 10$^{37}$ erg s$^{-1}$). Interestingly,
the latter luminosities from these Type IIL SNe are more 
consistent with a pulsar origin in young
SNe, such as SN~1968D, SN~1941C, and SN~1959D \citep{perna08,soria08}, 
in line with \citet{heger03} who point out that not all Type IIL's 
will form a black hole remnant.

\section{Conclusions}

Our analysis of archival X-ray observations of SN~1979C indicate that
the X-ray luminosity has been remarkably steady. We find that the
X-ray light curve is not consistent with either a model for a
supernova powered by a magnetar or a model where the X-ray emission
arises from a blast wave 
expanding into a dense circumstellar wind. In the latter case, the
observed decline in the optical and radio bands should be accompanied
by a decline in the X-ray flux. The X-ray spectrum for SN~1979C 
can be modeled by a combination of a thermal X-ray component and emission
from an accreting black hole with a mass $\sim$ 5M$_{\odot}$. The
accreting material likely originates from either a fallback disk 
after the supernova, or possibly from material accreted from a
binary companion which is suggested by the radio light curve
at early times \citep{weiler92, schwarz96}.  Finally, we note that 
the formation of a black
hole in SN~1979C might have also imprinted a feature in the optical
supernova light curve.  \citet{Young} recently 
showed that the supernova light curve can be fit by a two component
model which includes a GRB afterglow followed by supernova ejecta. 
They argued that
the GRB optical afterglow is produced when a jet, from the formation
of the central $\sim$ 2$M_{\sun}$ black hole, 
penetrates through the stellar envelope.

We note that SN~1979C appears to be the first example of a historic
supernova where there is possible evidence for a black hole remnant. A 
survey of Type IIL SNe observed at late times could reveal the existence
of other accreting black hole remnants and constrain the statistics
of black hole formation in core--collapse SNe. We note that a 
deep {\it Chandra} or {\it XMM--Newton} 
observation of SN~1979C could settle the issue, as
it would allow for both a detailed spectral analysis of the emitted
spectrum as well as a search for short term variations in the X-ray
light curve, which could be seen as evidence for ongoing accretion.

\bigskip
\paragraph*{Acknowledgments.}
We thank Roger Chevalier and Robert Fesen for useful comments on
an early draft of this manuscript.
This work was supported in part by NSF grant AST-0907890 and NASA
grants NNX08AL43G and NNA09DB30A for AL. DJP and CJ acknowledge
support from NASA Contract NAS8-03060 and the Smithsonian Institution.

\begin{deluxetable}{rrrrl}
\tablecolumns{5}
\tablewidth{0pc}
\tablecaption{X-ray observations of SN 1979C}
\tablehead{
\colhead{$\Delta t$} & \colhead{Count Rate} & 
\colhead{F$_X$\tablenotemark{a}} & \colhead{L$_X$\tablenotemark{b}} & 
\colhead{Mission} \\
\colhead{yr} & \colhead{10$^{-4}$ cps} & 
\colhead{10$^{-14}$ erg cm$^{-2}$ s$^{-1}$} & 
\colhead{10$^{38}$ erg s$^{-1}$} & \colhead{}}
\startdata
 0.7 & $<$ 3.0      & $<$ 2.3 & $<$ 6.3 & {\em Einstein} (HRI)\\
16.2  & 6.7$\pm$0.7  & 3.0$\pm$0.3 & 8.2$\pm$0.9 & {\em ROSAT} (HRI) \\
20.6  & 42.$\pm$2.0  & 2.5$\pm$0.2 & 6.9$\pm$0.6 & {\em Chandra} (ACIS-S) \\
22.7\tablenotemark{c}  & \nodata      & 2.3$\pm$0.3 & 6.3$\pm$0.7 & {\em XMM--Newton} (MOS) \\ 
26.5\tablenotemark{d} & 8.0$\pm$0.9  & 2.3$\pm$0.3 & 6.3$\pm$0.7 & {\em Swift} (XRT) \\
26.9  & 40.$\pm$0.8  & 2.4$\pm$0.2 & 6.6$\pm$0.5 & {\em Chandra} (ACIS-S) \\
28.0  & 43.$\pm$0.3  & 2.6$\pm$0.2 & 7.0$\pm$0.5 & {\em Chandra} (ACIS-S) \\
\enddata
\tablenotetext{a}{Unabsorbed 0.3--2.0 keV flux assuming a 0.5 keV
thermal Bremsstrahlung model with Galactic hydrogen column density 
of $2.4\times10^{20}$cm$^{-2}$.}
\tablenotetext{b}{Luminosity assumes a distance to M100 of 15.2$\pm$0.1 Mpc
\citep{freedman01}.}
\tablenotetext{c}{Flux value taken from \citet{immler05}.}
\tablenotetext{d}{Values are from {\em Swift} XRT observations taken
between 2005 and 2006, with the first observation performed on 2005--11--13.}
\label{tab:79c}
\end{deluxetable}

%\begin{table*}
%\centering
%\begin{minipage}{140mm}
%\caption{X-ray observations of SN~1979C}
%\begin{tabular}{@{}rrrrl@{}}
%\hline
%$\Delta t$ & Count rate & F$_X$\footnote{Unabsorbed 0.3--2.0 keV flux assuming a 0.5 keV thermal Bremsstrahlung model with Galactic hydrogen column density of $2.4\times10^{20}$cm$^{-2}$.} & L$_X$\footnote{Luminosity assumes a distance to M100 of 15.2$\pm$0.1 Mpc \citep{freedman01}.} & Mission \\
%yr        & 10$^{-4}$ cps & 10$^{-14}$ erg cm$^{-2}$ s$^{-1}$ & 10$^{38}$ erg & \\
%\hline
% 0.7 & $<$ 3.0      & $<$ 2.3 & $<$ 6.3 & {\em Einstein} (HRI)\\
%16.2  & 6.7$\pm$0.7  & 3.0$\pm$0.3 & 8.2$\pm$0.9 & {\em ROSAT} (HRI) \\
%20.6  & 42.$\pm$2.0  & 2.5$\pm$0.2 & 6.9$\pm$0.6 & {\em Chandra} (ACIS-S) \\
%22.7\footnote{Flux value taken from \citet{immler05}.} & \nodata     & 2.3$\pm$0.3 & 6.3$\pm$0.7 & {\em XMM--Newton} (MOS) \\ 
%26.9 & 40.$\pm$0.8  & 2.4$\pm$0.2 & 6.6$\pm$0.5 & {\em Chandra} (ACIS-S) \\
%29.0  & 43.$\pm$0.3  & 2.6$\pm$0.2 & 7.0$\pm$0.5 & {\em Chandra} (ACIS-S) \\
%\hline
%\end{tabular}
%\label{tab:79c}
%\end{minipage}
%\end{table*}

\begin{thebibliography}{}



\bibitem[Balberg et al.(2000)]{balberg00} Balberg, S., Zampieri,
L., \& Shapiro, S.~L.\ 2000, ApJ, 541, 860

\bibitem[Bartel 
\& Bietenholz(2003)]{bartel03} Bartel, N., \& Bietenholz, M.~F.\ 2003, ApJ, 591, 301 

\bibitem[Bartel 
\& Bietenholz(2008)]{bartel08} Bartel, N., \& Bietenholz, M.~F.\ 2008, ApJ, 682, 1065 

\bibitem[Bietenholz et al.(2010)]{bietenholz10} Bietenholz, M.~F., 
Bartel, N., \& Rupen, M.~P.\ 2010, ApJ, 712, 1057 

\bibitem[Chandra et al.(2009)]{chandra09} Chandra, P., Dwarkadas, 
V.~V., Ray, A., Immler, S., \& Pooley, D.\ 2009, \apj, 699, 388 

\bibitem[Chevalier \& Fransson(1994)]{Chev} Chevalier, R.  \&
Fransson, C. 1994. ApJ, 420, 268

\bibitem[Davis et al.(2006)]{Davis06} Davis, S.~W., Done, C., 
\& Blaes, O.~M.\ 2006, ApJ, 647, 525 

\bibitem[Dickey \& Lockman(1990)]{dickey90} Dickey, J.~M., \& 
Lockman, F.~J.\ 1990, ARA\&A, 28, 215 

\bibitem[Duncan 
\& Thompson(1992)]{duncan92} Duncan, R.~C., \& Thompson, C.\ 1992, 
ApJ, 392, L9 

\bibitem[Fesen 
\& Matonick(1993)]{fesen93} Fesen, R.~A., \& Matonick, D.~M.\ 1993, ApJ, 407, 110 

\bibitem[Freedman et al.(2001)]{freedman01} Freedman, W.~L., et 
al.\ 2001, ApJ, 553, 47 

\bibitem[Fryer 
\& Kalogera(2001)]{fryer01} Fryer, C.~L., \& Kalogera, V.\ 2001, ApJ, 554, 548 

\bibitem[Heger et al.(2003)]{heger03} Heger, A., Fryer, C.~L., 
Woosley, S.~E., Langer, N., \& Hartmann, D.~H.\ 2003, ApJ, 591, 288 

\bibitem[Immler et 
al.(1998)]{immler98} Immler, S., Pietsch, W., \& Aschenbach, B.\ 1998, A\&A, 331, 601 

\bibitem[Immler et al.(2002)]{immler02} Immler, S., Wilson, 
A.~S., \& Terashima, Y.\ 2002, \apjl, 573, L27 

\bibitem[Immler et al.(2005)]{immler05} Immler, S., et al.\ 
2005, ApJ, 632, 283 

\bibitem[Kaaret(2001)]{kaaret01} Kaaret, P.\ 2001, ApJ, 560, 
715 

\bibitem[Kasen 
\& Bildsten(2009)]{kasen09} Kasen, D., \& Bildsten, L.\ 2009, arXiv:0911.0680 

\bibitem[Kouveliotou et al.(1998)]{kouveliotou98} Kouveliotou, C., et 
al.\ 1998, Nature, 393, 235 

\bibitem[Marcaide et 
al.(2009)]{marcaide09} Marcaide, J.~M., Mart{\'{\i}}-Vidal, I., Perez-Torres, M.~A., Alberdi, A., Guirado, J.~C., Ros, E., \& Weiler, K.~W.\ 2009, A\&A, 503, 869 

\bibitem[Mattei et al.(1979)]{mattei79} Mattei, J., Johnson, 
G.~E., Rosino, L., Rafanelli, P., \& Kirshner, R.\ 1979, IAU Circ., 3348, 1 

\bibitem[McClintock et al.(2009)]{Mc} McClintock, J. E., et al. 2009,
preprint arXiv:0911.5408

\bibitem[Milisavljevic et al.(2009)]{milisavljevic09} Milisavljevic, 
D., Fesen, R.~A., Kirshner, R.~P., \& Challis, P.\ 2009, ApJ, 692, 839 

\bibitem[Montes et al.(2000)]{montes00} Montes, M.~J., Weiler, 
K.~W., Van Dyk, S.~D., Panagia, N., Lacey, C.~K., Sramek, R.~A., 
\& Park, R.\ 2000, ApJ, 532, 1124 

\bibitem[Perna et al.(2000)]{P00} Perna, R., Hernquist, L., \&
Narayan, R.\ 2000, ApJ, 541, 344

\bibitem[Perna et al.(2008)]{perna08} Perna, R., Soria, R., 
Pooley, D., \& Stella, L.\ 2008, MNRAS, 384, 1638 

\bibitem[Pindao et 
al.(2002)]{pindao02} Pindao, M., Schaerer, D., Gonz{\'a}lez Delgado, R.~M., 
\& Stasi{\'n}ska, G.\ 2002, A\&A, 394, 443 

\bibitem[Richardson et al.(2002)]{richardson02} Richardson, D., 
Branch, D., Casebeer, D., Millard, J., Thomas, R.~C., 
\& Baron, E.\ 2002, AJ, 123, 745 

\bibitem[Schwarz 
\& Pringle(1996)]{schwarz96} Schwarz, D.~H., \& Pringle, J.~E.\ 1996, 
MNRAS, 282, 1018 

\bibitem[Smartt(2009)]{smartt09} Smartt, S.~J.\ 2009, \araa, 47, 63 

\bibitem[Soria 
\& Perna(2008)]{soria08} Soria, R., \& Perna, R.\ 2008, ApJ, 683, 767 

\bibitem[Truelove 
\& McKee(1999)]{truelove99} Truelove, J.~K., \& McKee, C.~F.\ 
1999, ApJS, 120, 299 

\bibitem[van Dyk et al.(1999)]{vandyk99} van Dyk, S.~D., et al.\ 
1999, PASP, 111, 313 

\bibitem[Weiler et al.(1986)]{weiler86} Weiler, K.~W., Sramek, 
R.~A., Panagia, N., van der Hulst, J.~M., 
\& Salvati, M.\ 1986, ApJ, 301, 790 

\bibitem[Wang et al.(2006)]{Wang06} Wang, Z., Chakrabarty, D., \&
Kaplan, D.~L.\ 2006, Nature, 440, 772

\bibitem[Weiler et al.(1992)]{weiler92} Weiler, K.~W., van Dyk, 
S.~D., Pringle, J.~E., \& Panagia, N.\ 1992, ApJ, 399, 672 

\bibitem[Woosley(2009)]{woosley09} Woosley, S.~E.\ 2009, 
arXiv:0911.0698 

\bibitem[Young, Smith, \&Johnson(2005)]{Young}
Young, T.~R., Smith, D., \& Johnson, T.~A.\ 2005, ApJ, 625, L87

\bibitem[Zampieri et al.(1998)]{zampieri98} Zampieri, L., Colpi,
M., Shapiro, S.~L., \& Wasserman, I.\ 1998, ApJ, 505, 876



\end{thebibliography}
\end{document}